\documentclass[a4j,10pt,onecolumn,oneside,notitlepage,final]{article}
\usepackage{ amsmath }
\usepackage{ amssymb }
\usepackage{amsfonts}
\usepackage{amsthm}
\usepackage{amscd}
\usepackage{ latexsym }
\theoremstyle{plain}
\newtheorem{thm}{Theorem}[section]
\newtheorem{prop}[thm]{Proposition}
\newtheorem{cor}[thm]{Corollary}
\newtheorem{lem}[thm]{Lemma}

\theoremstyle{definition}

\newtheorem*{acknowledge}{Acknowledgments}

\begin{document}
\title{Non stabilizer Clifford codes with qupits}
\author{
  %˜a•¶'Ì'æˆê'˜ŽÒ–¼
  HAGIWARA Manabu
  \thanks{ %˜a•¶'̏Š'®'ƏZŠ
  %‰p•¶'̏Š'®'ƏZŠ
  Institute of Industrial Science, University of Tokyo,
  4-6-1 Komaba, Meguro-ku Tokyo, Japan
  % E-mail address
E-mail: {\tt \{manau, imai\}@\allowbreak
\{imailab.\allowbreak
iis, iis\}.\allowbreak
u-tokyo.\allowbreak
ac.\allowbreak
jp}
}\\
  %‰p•¶'Ì'æˆê'˜ŽÒ–¼
\and
 Hideki IMAI $^{*}$
}
\date{}
\maketitle
\abstract{
We present a method to construct a non stabilizer Clifford code which encodes a single qupit, i.e.  a state described as a vector in $p$-dimensional Hilbert space, to a pair of a single qupit and a single qubit, for any odd prime $p$.
Thus we obtain infinite non stabilizer Clifford codes.
}

\section{Introduction}
There are well-known classes of quantum error correcting codes, such as CSS codes, stabilizer codes, and Clifford codes.
Stabilizer codes can be seen as a generalization of a class of CSS codes.
In the same way, Clifford codes can be understood as a generalization of stabilizer codes.
To show the existence of a true Clifford code which is better than any stabilizer code is a well known open problem in the theory of Clifford codes.
One of the main difficulties in solving this problem is that we know only about 110 examples of codes which are Clifford but not stabilizer codes.
In this abstract, we obtain infinite examples of Clifford codes which are not stabilizer codes.
We expect our examples to be useful in the study of Clifford codes.

A Clifford code is constructed from a quadruple of parameters $(G, \rho,  N, \chi)$,  a finite group $G$, an irreducible, faithful unitary representation $\rho$ of $G$ with degree root of $(G: Z(G))$, a normal subgroup $N$, and an irreducible character $\chi$ of $N$.
Let $\sigma$ be a standard cyclic permutation matrix with degree $p$.
Let $\lambda$ be a primitive $p$-th root of $1$ and let $i$ be a primitive $4$-th root of $1$, that is, $i^{2}=-1$.
Let denote a diagonal matrix diag$(1, \lambda, \lambda^{2}, \dots , \lambda^{p-1} )$ by $\tau$.
Put $2p \times 2p $ matrices $A:=$ diag$(\sigma, \sigma^{-1})$, $B:=$ diag$(i \tau, i^{-1} \tau^{-1})$, and $C:= \left(
  \begin{array}{cc}
       &  I_{p}  \\
  -I_{p}     &    \\
  \end{array}
\right) $ where $I_{p}$ is the identity matrix.

Let $G$ be a group generated by $A, B,$ and $C$ and let $\rho$ be a matrix representation.
We show that the order of $G$ is $2^3 p^3$ and that the order of its center $Z(G)$ is $2p$ in \S\ref{orderG}.
Thus we have deg$\rho$=$(G : Z(G))^{1/2}$.
Let $N$ be a subgroup generated by $A, B^{4}$, and $C$. Then $N$ is shown to be normal and we have a $p$-dimensional $N$-invariant space $V$ in \S\ref{orderN}. 
Thus $V$ introduces a character $\chi$.
$\chi$ is shown to be irreducible.
Hence we have a Clifford code $V$ with its parameters $(G, \rho, N, \chi)$.
Moreover, we show that this code is not a stabilizer code in \S\ref{nonstab}.

\section{Notations, Definitions, and Preliminaries}
Let $G$ be a finite group having an irreducible, faithful unitary representation $\rho$ of degree $(G: Z(G))^{1/2}$ where $Z(G)$ is the center of $G$, and let $\phi$ be a character with respect to $\rho$.
Let $N$ be a normal subgroup of $G$, and let $\chi$ be an irreducible character of $N$ such that $( \chi , \phi_{N} ) \neq 0$, where $(,)$ is the inner product on characters.
A \textbf{Clifford code} with parameters $(G, \rho, N , \chi)$ is a representation with respect to $\chi$.
If the normal subgroup $N$ is abelian, then the Clifford code is called a \textbf{stabilizer code}.
If $N$ is not abelian, then the representation might still be a stabilizer code, but with respect to another normal abelian subgroup of $G$.
In \cite{KL2}, A. Klappenecker, M R\"{o}tteler showed the following.
\begin{thm}[\cite{KL2} Cor. 6]\label{thmKL}
Assume $N$ contains $Z(G)$.
A Clifford code $V$ with $(G, \rho, N, \chi)$ is a stabilizer code if and only if $\deg \rho = |N|/|H|$ for some abelian subgroup $H$ such that $H$ contains $Z(G)$ and $H$ is a subgroup of \textbf{quasikernel} on $\chi$ of $G$, i.e. any element of $H$ acts on $V$ as a scalar mapping.
\end{thm}

Let $p$ be an odd prime.
Let $\lambda$ be a primitive $p$-th root of $1$ and let $i$ be a primitive $4$-th root of $1$.
Let $\omega $ be a primitive $4p$-th root of $1$.
Let $\sigma$ be a standard cyclic permutation matrix with degree $p$, i.e.
$$ \sigma = \left(
  \begin{array}{ccccc}
    0   &     &    &    & 1   \\
    1   &  0  &    &    &    \\
        &  1  & \ddots   &   &    \\
        &     & 1  & 0  &    \\
    0   &     &    & 1   & 0   \\
  \end{array}
\right).
$$
Let $\tau$ be a matrix with degree $p$ such that
$$ \tau = \left(
  \begin{array}{ccccc}
    1   &    &    &    &    \\
       & \lambda   &    &    &    \\
       &    & \lambda^{2}   &    &    \\
       &    &    & \ddots   &    \\
       &    &    &    & \lambda^{p-1}   \\
  \end{array}
\right).
$$
Let $I_{p}$ be the identity matrix with degree $p$.

Put three matrices $A, B,$ and $C$ with degree $2p$ by the following,
$$A := \left(
  \begin{array}{cc}
    \sigma   &    \\
       &  \sigma^{-1}  \\
  \end{array}
\right), B := \left(
  \begin{array}{cc}
    i \tau   &    \\
       & i^{-1} \tau^{-1}   \\
  \end{array}
\right),
C := \left(
  \begin{array}{cc}
       &  I_{p}  \\
  -I_{p}     &    \\
  \end{array}
\right).
$$

\section{The order and the center of $G$}\label{orderG}
By the definition of $A$ and $B$, we have the following.
\begin{lem}\label{conjAB}
$$ABA^{-1} = \lambda B$$
\end{lem}

\begin{lem}\label{subG}
$\langle A , B \rangle = \{\lambda^{a}
\left(
  \begin{array}{cc}
    i^{b} \tau^{b} \sigma^{c}   &    \\
       & i^{-b } \tau^{-b} \sigma^{-c}   \\
  \end{array} 
\right) 
| 0 \le a < p , 0 \le b < 4p, 0 \le c < p \}$.
In particular, we have $\# \langle A , B \rangle = 2^{2} p^{3}$
\end{lem}
\begin{proof}
By lemma \ref{conjAB}, we have $\langle A , B \rangle = \langle A , B , \lambda I_{2p}\rangle$.
Since $A,B$ are diagonal matrices with 2 blocks of size $p$, we have $\langle A , B \rangle \simeq \langle \sigma , i \tau , \lambda I_{p}\rangle$, where $\simeq$ means isomorphic as a group.

$\langle \sigma , i \tau , \lambda I_{p}\rangle$ has a normal subgroup $\langle i \tau, \lambda I_{p}\rangle$.
Furthermore $\langle \sigma , i \tau , \lambda I_{p}\rangle$ is a semi-product of $\sigma$ and $\langle i \tau, \lambda I_{p}\rangle$.

The structure of $\langle i \tau, \lambda I_{p}\rangle$ is easily obtained as follows,
$$\langle i \tau , \lambda I_{p}\rangle = \{ \lambda^{a} (i \tau)^{b} | 0 \le a < p , 0 \le b < 4p\} = \{ \omega^{a} \tau^{b} | 0 \le a < 4p , 0 \le b < p\} .$$

On the other hand, the order of $\sigma$ is $p$.
Hence we have 
$$ \# \langle A , B \rangle = 2^{2} p^{3}.$$

Furthermore, we have 
$$ \langle A , B \rangle = \{
\lambda^{a}
\left(
  \begin{array}{cc}
    (i \tau)^{b} \sigma^{c}   &    \\
       & (i \tau)^{-b} \sigma^{-c}   \\
  \end{array} 
\right) 
| 0 \le a < p , 0 \le b < 4p, 0 \le c < p \}
$$
\end{proof}

\begin{prop}\label{stG}
$ G = 
\{
\lambda^{a}
\left(
  \begin{array}{cc}
    (i \tau)^{b} \sigma^{c}   &    \\
       & (i \tau)^{-b} \sigma^{-c}   \\
  \end{array} 
\right) 
| 0 \le a < p , 0 \le b < 4p, 0 \le c < p \}
 \cup$$
$$ 
\{
\lambda^{a}
\left(
  \begin{array}{cc}
   &  (i \tau)^{b} \sigma^{c}    \\
   -(i \tau)^{-b} \sigma^{-c} &  \\
  \end{array} 
\right) 
| 0 \le a < p , 0 \le b < 4p, 0 \le c < p \}
.
$
In particular, we have $\# G = 2^{3} p^{3}$
\end{prop}
\begin{proof}
Now we have $C^{2} = -I_{2p}$.
By lemma \ref{subG}, we can describe the elements of $G$ as follows,
$$ G = 
\{
\lambda^{a}
\left(
  \begin{array}{cc}
    (i \tau)^{b} \sigma^{c}   &    \\
       & (i \tau)^{-b} \sigma^{-c}   \\
  \end{array} 
\right) 
| 0 \le a < p , 0 \le b < 4p, 0 \le c < p \}
 \cup$$
$$ 
\{
\lambda^{a}
\left(
  \begin{array}{cc}
   &  (i \tau)^{b} \sigma^{c}    \\
   -(i \tau)^{-b} \sigma^{-c} &  \\
  \end{array} 
\right) 
| 0 \le a < p , 0 \le b < 4p, 0 \le c < p \}
.
$$
\end{proof}
Let $\rho$ be the matrix representation of $G$, in other words, $\rho(g) = g$ for any $g \in G$.
\begin{cor}\label{irrG}
$\rho$ is irreducible.
\end{cor}
\begin{proof}
$$
\frac{1}{|G|} \sum_{g \in G} | \mathrm{tr}(g) |^{2}
= \frac{1}{2^{3} p^{3}} \sum_{0 \le a < p} |2p \lambda^{a}|^{2} = 1
$$
\end{proof}
Let denote the center of $G$ by $Z(G)$.
\begin{cor}
$\# Z(G) = 2p$
\end{cor}
\begin{proof}
Since $\rho$ is irreducible, $Z(G)$ consists of the scolor mappings of $G$.
By proposition \ref{stG}, we have $\# Z(G) = 2p$.
\end{proof}
\begin{cor}
$\deg \rho = ( G : Z(G) )^{1/2}$ 
\end{cor}
\begin{proof}
We have $\deg \rho = 2p$ and $(G : Z(G) )^{1/2} = (2^{3} p^{3} / 2p)^{1/2} = 2p$.
\end{proof}

\section{properties of $N$}\label{orderN}
\begin{prop}\label{stN}
$N = \{ 
\omega^{2 a}
\left(
  \begin{array}{cc}
    \tau^{b} \sigma^{c}   &    \\
       & \tau^{-b} \sigma^{-c}   \\
  \end{array} 
\right),
\omega^{2 a}
\left(
  \begin{array}{cc}
  &  \tau^{b} \sigma^{c}     \\
  - \tau^{-b} \sigma^{-c}  &   \\
  \end{array} 
\right),
| 0 \le a < 2p , 0 \le b,c < p \}
$.
In particular, we have $\# N = 2^{2} p^{3}$.
\end{prop}
\begin{proof}
We note that $\langle A , B^{4} \rangle = \langle A , B^{4} , \lambda I_{2p} \rangle$.
It is easy to show that 
$$\langle A , B^{4} , \lambda I_{2p} \rangle \simeq \langle \sigma , \tau^{4} , \lambda I_{p} \rangle.$$

By an argument similar to that of  proposition \ref{stG}, we have
$$\# \langle A , B^{4} \rangle = p^{3}.$$
Furthermore, we have
$$\langle A , B^{4} \rangle = \{ 
\lambda^{a}
\left(
  \begin{array}{cc}
    \tau^{b} \sigma^{c}   &    \\
       & \tau^{-b} \sigma^{-c}   \\
  \end{array} 
\right) 
| 0 \le a,b,c < p \}.
$$

Hence it follows
$N = \langle A , B^{4} , C \rangle$
$$ = \{ 
\omega^{2 a}
\left(
  \begin{array}{cc}
    \tau^{b} \sigma^{c}   &    \\
       & \tau^{-b} \sigma^{-c}   \\
  \end{array} 
\right),
\omega^{2 a}
\left(
  \begin{array}{cc}
  &  \tau^{b} \sigma^{c}     \\
  - \tau^{-b} \sigma^{-c}  &   \\
  \end{array} 
\right),
| 0 \le a < 2p , 0 \le b,c < p \}.
$$
\end{proof}
\begin{cor}
$N$ is a normal subgroup of $G$ and contains $Z(G)$.
\end{cor}
\begin{proof}
Since $(G:N) = 2$, $N$ is a normal subgroup.
By proposition \ref{stN}, $N$ contains $Z(G)$.
\end{proof}

Let denote the standard basis of $\mathbb{C}^{2p}$ by $\{e_{h} | 1 \le h \le 2p\}$ and put
$$ V_{1} := \langle e_{\mathbf{c}^{x}(1)} -i e_{\mathbf{c}^{-x}(1)+p} \rangle_{ 0 \le x < p },$$
$$ V_{2} := \langle e_{\mathbf{c}^{x}(1)} +i e_{\mathbf{c}^{-x}(1)+p} \rangle_{ 0 \le x < p }.$$

We denote a cyclic permutation $(1,2, \dots, p)$ by $\mathbf{c}$, i.e. $\mathbf{c}(h) = h+1 \pmod{p}$ for any $1 \le h \le p$.
\begin{prop}\label{v12_inv}
$V_{1}, V_{2}$ are $N$-invariant.
\end{prop}
\begin{proof}
First, we show that $V_{1}$ is $N$-invariant.
We have that
$$A (e_{\mathbf{c}^{x}(1)} -i e_{\mathbf{c}^{-x}(1)+p}) = e_{\mathbf{c}^{x+1}(1)} -i e_{\mathbf{c}^{-(x+1)}(1)+p},$$
$$B^{4} (e_{\mathbf{c}^{x}(1)} -i e_{\mathbf{c}^{-x}(1)+p}) = \lambda^{4x}(e_{\mathbf{c}^{x}(1)} -i e_{\mathbf{c}^{-x}(1)+p}),$$
$$C (e_{\mathbf{c}^{x}(1)} -i e_{\mathbf{c}^{-x}(1)+p}) = -i (e_{\mathbf{c}^{-x}(1)} -i e_{\mathbf{c}^{x}(1)+p}),$$
for any $1 \le x \le p$.
Thus $V_{1}$ is $N$-invariant.

By a similar argument, $V_{2}$ can be proved to be $N$-invariant.
\end{proof}

Denote the character of $N$ of action on $V_{h}$ by $\chi_{h}$ for $h=1,2$.

\begin{prop}
$\chi_{1}$ and $\chi_{2}$ are irreducible as a character of $N$.
\end{prop}
\begin{proof}
By a similar caluculation to corollary \ref{irrG}, we have
$$\frac{1}{|N} \sum_{n \in N}|\mathrm{tr}(n)|^{2} = 2.$$
It implies that $\rho_{|N}$ consists of two irreducible characters where $\rho_{N}$ is the restrected representation of $\rho$ to $N$.
By proposition \ref{v12_inv}, we know two representation subspace $V_{1}, V_{2}$.
It is easy to verify $V_{1} \cap V_{2} = \{ 0 \}$.
Thus $\chi_{1}, \chi_{2}$ are ireducible.
\end{proof}

Therefore we have the following.
\begin{thm}
$V_{1}$ is a Clifford code with its parameter $(G, \rho, N, \chi)$.
\end{thm}

\section{non-stabilizer property}\label{nonstab}
\begin{prop}
The quasikernel on $\chi$ of $G$ is $Z(G)$.
\end{prop}
\begin{proof}
Let $0 \le a < 2p, 0 \le b, c < p$.
Put $X :=  \omega^{2 a}
\left(
  \begin{array}{cc}
    (i \tau)^{b} \sigma^{c}  &    \\
       & (i \tau)^{-b} \sigma^{-c} \\
  \end{array} 
\right)$.
It is easy to verify that $X$ is not a quasikernel if $k \neq 0$.

Thus we assume $c = 0$.
Then we have
$$ X (e_{\mathbf{c}^{x}(1)} -i e_{\mathbf{c}^{-x}(1)+p})
$$
$$
= \omega^{2a + 4 (x-1)b}
\left(
   \begin{array}{cc}
     i^{b} &  \\
     &  i^{-b}\\
   \end{array}
\right)
( e_{\mathbf{c}^{x}(1)} -i e_{\mathbf{c}^{-x}(1)+p} )
$$
$$
= \omega^{2a + 4 (x-1)b}
( i^{b} e_{\mathbf{c}^{x}(1)} -i^{-b+1} e_{\mathbf{c}^{-x}(1)+p} )
$$
$$
= \omega^{2a + 4 (x-1)b + bp}
( e_{\mathbf{c}^{x}(1)} -i^{-2b+1} e_{\mathbf{c}^{-x}(1)+p} ).
$$
Thus $X$ is quasikernel if $b=0$.

Put $Y := \omega^{2 a}
\left(
  \begin{array}{cc}
  &  (i \tau)^{b}      \\
  - (i \tau)^{-b}   &   \\
  \end{array} 
\right).$
Then we have 
$$
Y (e_{\mathbf{c}^{x}(1)} -i e_{\mathbf{c}^{-x}(1)+p})
$$
$$
=\omega^{2 a}
\left(
  \begin{array}{cc}
  (i \tau)^{b} &     \\
  & (i \tau)^{-b} \\
  \end{array} 
\right)
\left(
  \begin{array}{cc}
  & I_{p} \\
  -I_{p} & \\
  \end{array}
\right)
(e_{\mathbf{c}^{x}(1)} -i e_{\mathbf{c}^{-x}(1)+p})
$$
$$
=\omega^{2 a}
\left(
  \begin{array}{cc}
  (i \tau)^{b} &     \\
  & (i \tau)^{-b} \\
  \end{array} 
\right)
(-e_{\mathbf{c}^{x}(1)+p} -i e_{\mathbf{c}^{-x}(1)})
$$
$$
=-i\omega^{2 a}
\left(
  \begin{array}{cc}
  (i \tau)^{b} &     \\
  & (i \tau)^{-b} \\
  \end{array} 
\right)
(e_{\mathbf{c}^{-x}(1)} -ie_{\mathbf{c}^{x}(1)+p})
$$
$$
=-i \omega^{2 a}
(\omega^{4 (-x-1)b} i^{b}  e_{\mathbf{c}^{-x}(1)} -i \omega^{-4 (-x-1) b} i^{b} e_{\mathbf{c}^{x}(1)+p}).
$$
Thus $Y$ is not an element of the quasikernel.

Therefore the quasikernel of $G$ is $Z(G)$.
\end{proof}

\begin{prop}
$V_{1}, V_{2}$ are non-stabilizer Clifford codes.
\end{prop}
\begin{proof}
A group $H$ which contains $Z(G)$ and which is contained by the quasikernel on $V_{i}$ of $G$ must be $Z(G)$ for $i=1,2$.
Thus we have 
$$ \chi(1)^{2} = p^{2} \neq 2^{2} p^{3} / 2p = 2p^{2} = |N|/|H|.$$
By Theorem \ref{thmKL}, this Clifford code is non-stabilizer.
\end{proof}
\begin{acknowledge}
This work was supported by the project on ``Research and Development on Quantum Cryptography'' of Telecommunications Advancement Organization as part of the programme ``Research and Development on Quantum Communication Technology'' of the Ministry of Public Management, Home Affairs, Posts and Telecommunications of Japan.
\end{acknowledge}

\end{document}